\documentstyle[aps,12pt]{revtex}

\begin{document}
\title{The formation of the non-magnetic state\\
for the d$^6$ electron system.}
\author{R.J. Radwa\'{n}ski}
\address{Center for Solid State Physics, \'{s}w. Filip 5, 31-150 Krak\'{o}w,\\
Inst. of Physics, Pedagogical University, 30-084 Krak\'{o}w, Poland.\\
email: sfradwan@cyf-kr.edu.pl}
\author{Z. Ropka}
\address{Center for Solid State Physics, \'{s}w. Filip 5, 31-150 Krak\'{o}w.}
\maketitle

\begin{abstract}
It has been shown that the crystal-field interactions can produce a
non-magnetic singlet ground state for the highly-correlated d$^6$ system
situated in the quasi-octahedral crystal field (CEF) provided the spin-orbit
coupling is correctly taking into account. The intra-atomic spin-orbit
coupling in combination with the trigonal distortion of the cubic CEF
interactions produce a peculiar energy level scheme with a non-magnetic
ground state and highly-magnetic excited states at 11 and 70 meV. The
calculated temperature dependence of the magnetic susceptibility exhibits
very substantial departure from the Curie law that is due to second-order
CEF interactions. The present considerations can be applied to Co$^{3+}$ and
Fe$^{2+}$-ion compounds. The derived electronic and magnetic properties
mimic very much those found experimentally for LaCoO$_3$.
\end{abstract}

\pacs{71.70.E 75.10.D 75.30.Gw}

{\bf 1. Introduction}

A proper consistent description of paramagnetic 3d ions is still under
debate [1-7], though Sir Nevill Mott already in 1949 has realized that it is
strong electron correlations that play the fundamental role in determination
of the electronic structure of 3d-ion compounds, named later Mott
insulators. The electronic structure governs electronic and magnetic
properties. Despite of almost 50 years enormous theoretical research
activity in this fiel Nature once by once shows phenomena that reveal a
large shortage of our general understanding of magnetism of 3d-ion
compounds. Surely properties like these exhibited by LaCoO$_3$ are
challenging. At first, it has a non-magnetic ground state at low
temperatures [1-6]. Such the state always draws attention of the magnetic
community appealing the question about the nature of the magnetic moment.
Secondly, this non-magnetic ground state seems to transform with increasing
temperature into a magnetic state as is inferred from an unusual temperature
dependence [6,8] of the magnetic susceptibility $\chi $ ($\chi \cdot $T
instead of being constant continuously grows up with temperature). These two
facts resemble somehow the Kondo problem for the formation of the
non-magnetic state of a magnetic impurity. This phenomenon for LaCoO$_3$ is
discussed, following Raccah and Goodenough [6], by consideration a model
with substantial temperature variation in the population of low- and
high-spin Co ions, the latter being randomly distributed in the majority of
low-spin Co ions at low temperatures.

Different theoretical approaches for the 3d-ion compounds differ in the
starting estimation of the strength of 3d electron-electron correlations
with respect to the strength of crystal-field (CEF) interactions. The
spin-orbit (s-o) coupling is usually ignored basing on the consensus that
the s-o coupling for 3d ions is the smallest among these above-mentioned
interactions. This consensus is a reason for the misleading, according to
us, neglection of the s-o coupling in evaluations of the electronic
structure. Our studies indicate that it is just opposite - the smaller s-o
coupling the lower available energy excitations and the lower temperatures
with anomalies of electronic properties. Direct calculations confirm that
just a weak s-o coupling causes dramatic change of the electronic structure
by producing the fine electronic structure with low-energy excitations even
so small as 1 meV (=11.6 K = 8.0 cm$^{-1}$).

The aim of this paper is to study the possibility for the formation of a
non-magnetic state on the atomic scale for the d$^6$ electronic system in
case of the predominantly cubic surrounding. This paper has been motivated
by an unsolved problem of the cause of the non-magnetic ground state of LaCoO%
$_3$ [1-8].

{\bf 2.Theoretical outline}

In the insulator LaCoO$_3$ the cobalt atoms occur in the trivalent state
anticipated from the charge neutrality La$^{3+}$Co$^{3+}$O$_3^{2-}$. The Co$%
^{3+}$ ion has 6 d-electrons in the incomplete outer shell and here they
will be treated as forming the highly-correlated electron system 3d$^6$.
Owing to the perovskite-like structure of LaCoO$_3$ the Co$^{3+}$ ions are
situated in the octahedral cubic surrounding that allows the trigonal
distortion easily to occur along the cube diagonal.

A modern approach to compound like LaCoO$_3$ is based on an idea of Mott
that it is strong electron correlations that make electrons in the
incomplete 3d shell to stay rather localized than itinerant (Mott
insulators). The physical situation of the 3d$^6$ system of a
3d-transition-metal ion is here taken to be accounted for by considering the
single-ion-like Hamiltonian containing the electron-electron interactions
within the 3d shell $H_{el-el}$, the crystal-field $H_{CF}$, spin-orbit $%
H_{s-o}$ and Zeeman $H_Z$ interactions:

\begin{center}
$H_d=H_{el-el}+H_{CF}+H_{s-o}+H_Z(1).$
\end{center}

The electron-electron and spin-orbit interactions are intra-atomic
interactions, whereas crystal-field and Zeeman interactions account for
interactions of the unfilled 3d shell with the charge and spin surrounding.
These interactions are written in the decreasing-strength succession. This
point of view is in contrast to most of standard band-structure calculations
that assume the dominancy of crystal-field interactions i.e. starts from the
strong crystal-field approach.

In a zero-order approximation the electron correlations are accounted for by
phenomenological Hund's rules that yield for the 3d$^6$ electron
configuration the term $^5$D with S =2 and L=2 to be the ground term.
Following the intermediate crystal-field approach [9-11] the ground term $^5$%
D, under the action of the crystal field of the cubic symmetry splits into
the orbital triplet $\Gamma _5$, denoted also as T$_{2g}$, and the orbital
doublet $\Gamma _3$ (E$_g$). The $^5$D term is 25-fold degenerated; 5-fold
degeneration occurs with respect to the orbital degree of freedom. Each
orbital state contains the 5 spin-degree of freedom. The removal of this
25-fold degeneracy can be traced by consideration of the single-ion-like
Hamiltonian written for the lowest Hund's rule $\left| \text{LS}%
\right\rangle $ term:

\begin{center}
$H_d=$B$_4(O_4^0+$5$O_4^4)+k\lambda _oL\cdot S+\mu _{\text{B}}(L+$g$%
_eS)\cdot B_{ext}(2).$
\end{center}

The first term is the cubic CEF Hamiltonian with the Stevens operators $%
O_n^m $ that depend on the orbital quantum numbers L and its z-component L$_z
$. The second term accounts for the spin-orbit interactions; the coefficient
k denotes the change of the s-o coupling in a solid compared to the free-ion
value, presumably due to the covalent mixing. The last term accounts for the
influence of the magnetic field, the externally applied in the present case.
g$_e$ value equals to 2.0023.

{\bf 3. Results and discusion}

The calculations of the many-electron states of the 3d$^6$ system have been
performed by the diagonalization of a 25$\times $25 matrix associated with
the Hamiltonian (1) considered in the $\left| \text{LSL}_z\text{S}%
_z\right\rangle $ base [12]. Fig. 1 presents a general overview of the
effect of CEF and s-o interactions on the $^5$D term for the d$^6$ system ($%
\lambda $%
\mbox{$<$}
0) in case of cubic CEF interactions of the octahedral surroundings (T$_{2g}$
ground cubic subterm).

As a result of the computation one obtains the energies of the 25 states and
the eigenvectors containing information e.g. about the magnetic properties.
These magnetic characteristics are computationally revealed under the action
of the external magnetic field, and they are shown in Fig. 2 on the right
hand side. Of course, due to the spin-orbit coupling the states are no
longer purely cubic orbital states. The calculations have been performed for
the spin-orbit constant $\lambda _o$ for the Co$^{3+}$ ion of -210 K taken
after ref. 9, p.399 and with k=3.0. Values for B$_4$ depend on the strength
of cubic CEF interactions. The separation $\Delta $ of the E$_g$ and T$_{2g}$
states amounts to 120 B$_4$. The experimentally derived values for $\Delta $
are largely inconsistent: 1.2 eV is given in Ref. 3 or 2.4 eV in Ref.4. Thus
we have chosen for calculations B$_4$=+200 K that yields more reasonable
value of $\Delta $ of 2.1 eV. The positive sign for B$_4$ is consistent with
the octahedral oxygen surrounding in LaCoO$_3$. It yields the orbital
triplet T$_{2g}$ cubic subterm as the ground state. It is 15-fold
degenerated in the LS space. The spin-orbit interactions largely removes the
15 or 10-fold degeneracy of the T$_{2g}$ and E$_g$ subterms as is seen in
Fig. 1d and 2b. As we are mostly interested in the ground state as it
determines properties at low temperatures we would like to point out that
the lowest triplet state, in the orbital+spin space, can be split by
electrostatic interactions as is shown in Fig. 2c. We have checked that a
trigonal or tetragonal distortion, described by the leading term B$_2^0O_2^0$%
, removes the three-fold degeneracy of the lowest cubic CEF+s-o state
producing the singlet and the doublet. The positive B$_2^0$ term along the
cube diagonal (the trigonal distortion) realizes the Jahn-Teller theorem,
i.e. makes the singlet to be energetically lower. This discusion is relevant
to LaCoO$_3$ as it undergoes a trigonal (rhombohedral) distortion at
temperature of 1210 K [6].

Fig. 2 presents the influence of the spin-orbit coupling and of the trigonal
distortion on the cubic CEF subterm $^5$T$_{2g}$. The resulting electronic
structure, shown in Fig. 2c, is quite peculiar. It has i) a non-magnetic
ground state and ii) the highly magnetic excited state; in fact even two
highly-magnetic doublets. The magnetic moment of the first and the second
excited states amounts to $\pm $2.32 $\mu _B$ and $\pm $3.66 $\mu _B$. The
Jahn-Teller splitting depends on the value of the trigonal distortion B$_2^0$%
. The calculated temperature dependence of the magnetic susceptibility is
shown in Fig.3a. It exhibits very special behaviour with a large maximum at
temperature of ~$\sim $ 90 K. In fact, the trigonal distortion B$_2^0$ of
+180 K and the 3-fold increase of the s-o coupling have been chosen in order
to have this maximum at 90 K. For smaller values B$_2^0$ and k this maximum
occurs for lower temperatures. Obviously, the special behaviour $\chi $(T)
results from the peculiarities of the local electronic structure. $\chi $
does not follow the Curie law even at higher temperatures what is clearly
seen in the $\chi ^{-1}$ vs T plot presented in Fig. 3a. It mimics a
Curie-Weiss (C-W) behaviour with a quite large value of the paramagnetic
Curie temperature $\theta $. $\theta $ is, however, largely
temperature-dependent varying from 270 K to 600 K. This substantial
departure from the Curie law, mimicing the C-W law, is purely the CEF effect
associated with the second-order CEF interactions. There is an apparent
change of the slope of the $\chi ^{-1}$(T) plot at 300 K that mimics somehow
the one found experimentally in LaCoO$_3$ [6]. An effective moment
calculated from this slope in the temperature interval 150-300 K yields 5.0 $%
\mu _{\text{B}}$ and 6.3 $\mu _{\text{B}}$ if calculated from the 400-1000 K
interval. This change of the effective moment is even more visible in the
plot $\chi \cdot $T vs T that is presented in Fig. 3b. It is the standard
plot for the presentation of the Curie low in the physical chemistry [13].
The departure from the constancy of $\chi \cdot $T has been taken as the
signal of the temperature variation of the effective moment like it was in
case of LaCoO$_3$ [8,6]. Using the Curie - law formula $p^{eff}$ =$\sqrt{%
\text{3}\chi \cdot \text{T}}$ one gets 0, 2.63, 3.56, 5.00 $\mu _{\text{B}}$
at 0, 100, 300 and 1000 K, respectively (see also Fig.3b). Values of 2.82
and 4.90 $\mu _{\text{B}}$ one finds from the theoretical spin-only
expression for S=1 and S=2, respectively. It means that this purely CEF
effect can be misleadingly attributed to a temperature-induced low-spin to
high-spin transformation. Finally it is worth noting that that the shape of
calculated results in Fig. 3b are in nice agreement with very detailed
experimental data presented in Ref. 13, Fig. 4.

{\bf 4. Conclusions}

It has been shown that the crystal-field interactions can produce a
non-magnetic singlet ground state for the highly-correlated d$^6$ system
situated in the quasi-octahedral crystal-field surrounding provided the
intra-atomic spin-orbit coupling is correctly taking into account [14]. The
s-o coupling produces for the d$^6$ system the fine electronic structure
with 15 available energy states below 0.3 eV. The spin-orbit interactions in
combination with the trigonal distortion of the cubic CEF interactions
produce a peculiar energy level scheme with a non-magnetic ground state and
highly-magnetic excited states at 11 meV and 70 meV. The calculated
temperature dependence of the magnetic susceptibility exhibits very
substantial departure from the Curie law. In such the case one can
misleadingly evaluate the effective paramagnetic moment and reveal its
strong temperature variation. The present considerations about the d$^6$
system can be applied to compounds containing Co$^{3+}$, Fe$^{2+}$, Rh$^{3+}$
and Ru$^{2+}$ ions. The derived electronic and magnetic properties mimic
very much those found experimentally in LaCoO$_3$. The natural explanation
for the insulating state of LaCoO$_3$, the evaluation of the fine electronic
structure and the possibility for calculations of the influence of
temperature on electronic and magnetic properties are great advantages of
the presented approach.

Fig. 1. a) The 25-fold degenerated $^5$D term of the 3d$^6$ system that is
expected to be realized in the Co$^{3+}$/Fe$^{2+}$ ion. b) The energy level
scheme for the 3d$^6$ system under the action of CEF interactions of the
octahedral symmetry. All the orbital levels have the internal 5-fold
spin-degree of freedom., c) the effect of the intra-atomic spin-orbit
coupling. interactions on the localized states of the 3d$^6$ system
according to Abragam and Bleaney, Fig. 7.19; c) the effect of the s-o
coupling on the localized states of the 3d$^6$ system according to the
first-order perturbation method, see e.g. Ref. 9 Fig. 7.19, the combined
effect of the cubic CEF\ and the s-o coupling according to the present
calculations: additional splittings revealed by the present exact
calculations should be noticed.

Fig. 2. The influence of the trigonal distortion, B$_2^0$=+180 K, on the
cubic CEF+spin-orbit states ($\lambda _o$=-210 K, k=3 B$_4$=+200 K) of the
15-fold degenerated orbital-triplet $^5$T$_{2g}$ ground state. The trigonal
distortion produces the fine electronic structure with a nonmagnetic singlet
ground state and strongly magnetic excited states at 125 K and 810 K (c).
The states are labelled with the degeneracy, the respective energy and the
magnetic moment.

Fig. 3. a) The calculated temperature dependence of the magnetic
susceptibility $\chi $ (the solid line) for the 3d$^6$ system situated in a
trigonally distorted cubic octahedral site. It does not follow the Curie law
having a sharp rounded peak at $\sim $~90 K. Two distingly different regions
in the $\chi ^{-1}$ vs T plot (the dot-dashed line) are seen: below 300 K
with $p^{eff}$=5.0 $\mu _{\text{B}}$ and above 400 K with $p^{eff}$ of 6.3 $%
\mu _{\text{B}}$. The $\chi ^{-1}$ vs T dependence mimics the Curie-Weiss
behaviour though this anomalous susceptibility dependence is purely CEF
effect associated with the substantial second-order CEF term. The
susceptibility calculated without the B$_2^0$ distortion (dotted line)
almost follows the Curie law with a mean value of $p^{eff}$=5.84 $\mu _{%
\text{B}}$. In the top the energy positions of the three lowest states are
shown. b) Temperature dependence of the $\chi \cdot $T (the solid line). A
continuous increase indicates departure from the Curie law. It could be
misleadingly attributed to the temperature-induced increase of the effective
paramagnetic moment, shown by the dashed line, if one misleadingly uses the
Curie-law expression $p^{eff}$= $\sqrt{\text{3}\chi \cdot \text{T}}$. The
general shape of $\chi $(T) is in nice agreement with experimental data
presented in Fig.4 of Ref. 13.


\begin{references}
\bibitem{12}  the computer program is available on the written request to
the author.
\end{references}
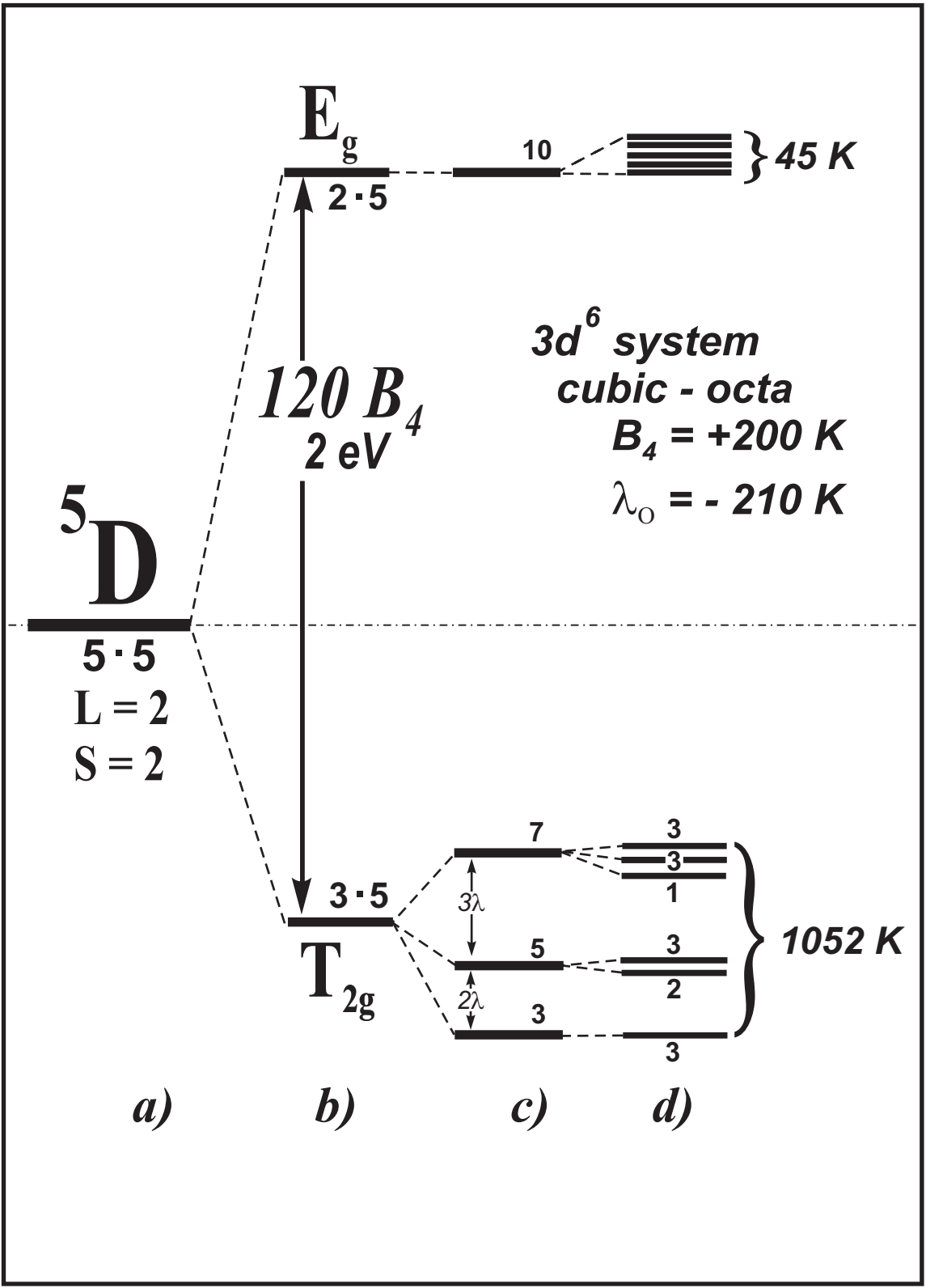
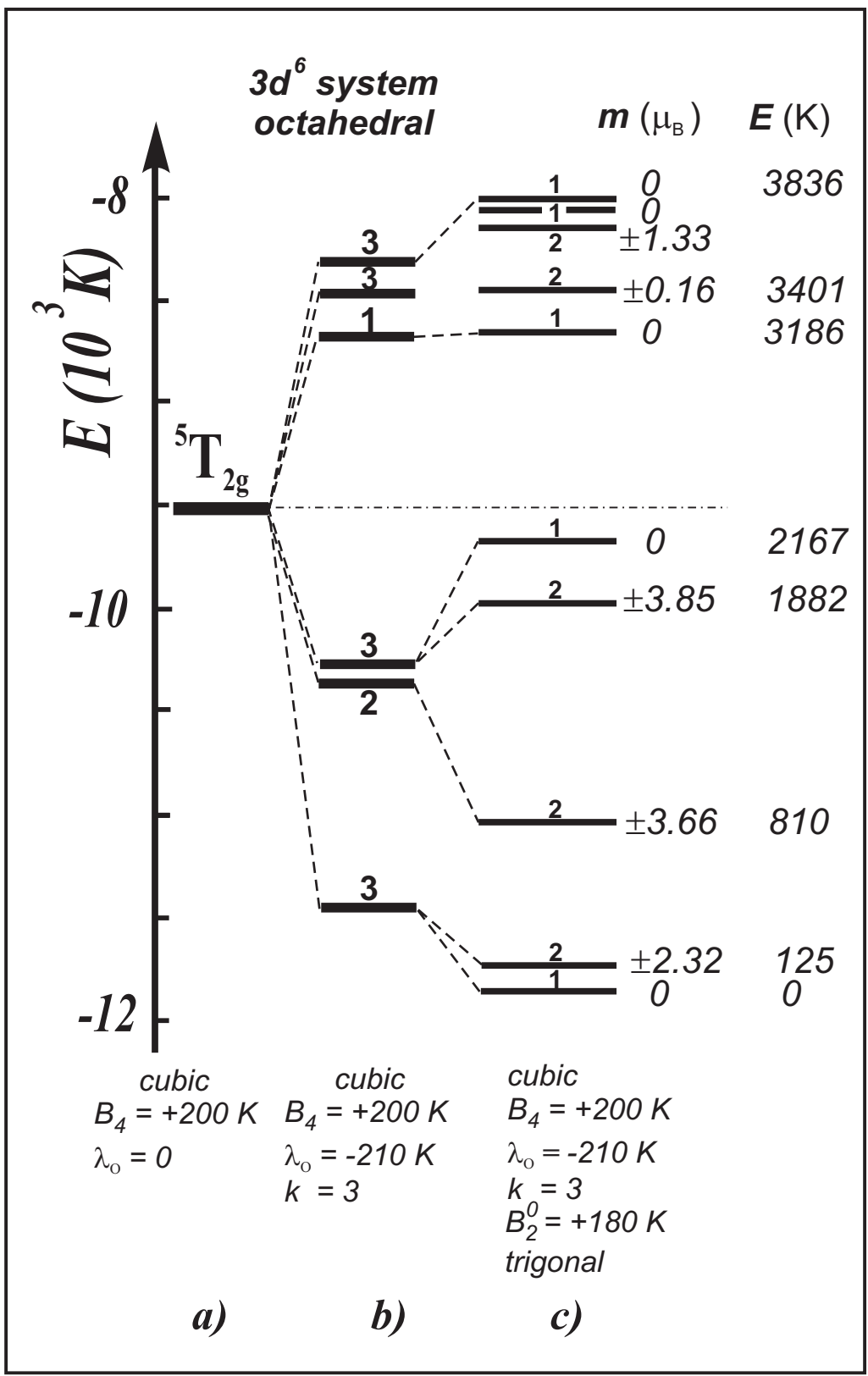
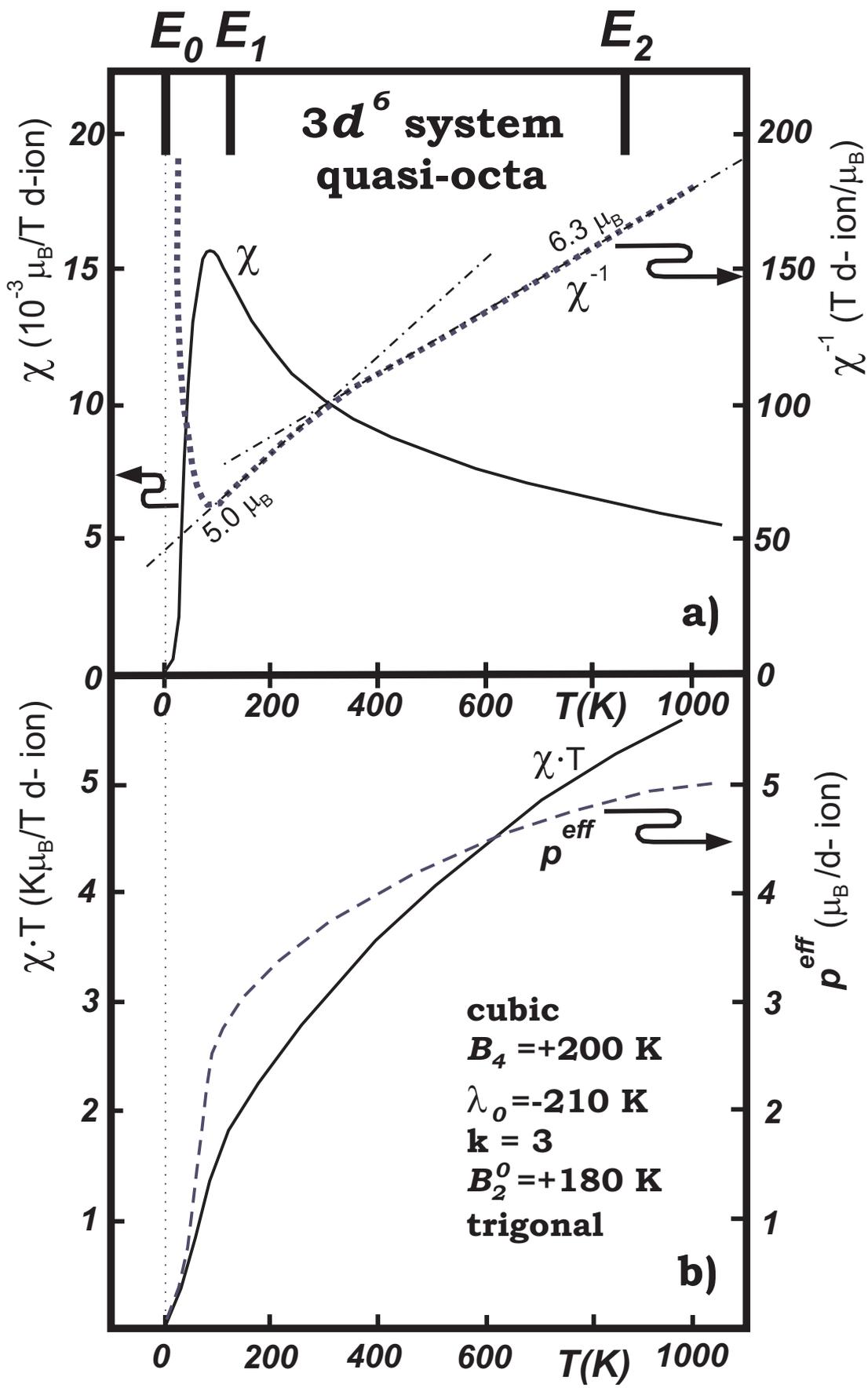
\end{document}